\documentstyle[aps,epsfig,floats]{revtex}
\newcommand{\bm}[1]{\mbox{\boldmath $#1$}}
\newcommand{\be}{\begin{equation}}
\newcommand{\ee}{\end{equation}}
\newcommand{\bea}{\begin{eqnarray}}
\newcommand{\eea}{\end{eqnarray}}
\newcommand{\st}{{\scriptscriptstyle T}}

\def\slash{\rlap{/}}

\begin{document}
\tighten
\thispagestyle{empty}
\title{
\begin{flushright}
\begin{minipage}{4 cm}
\small
hep-ph/9903354\\ 
VUTH 99-03
\end{minipage}
\end{flushright}
\vspace{3mm}
Time-reversal odd fragmentation and distribution functions\\
 in pp and ep single spin asymmetries.
\protect} 
\vspace{3mm}
\author {M. Boglione and P.J. Mulders\\  
\vspace{3mm}
\mbox{}\\
{\it Division of Physics and Astronomy, Faculty of Science, Free University}\\
{\it De Boelelaan 1081, NL-1081 HV Amsterdam, the Netherlands}}

\maketitle

\vspace{1cm}

\begin{abstract}
We present some estimates of T-odd fragmentation and distribution functions, 
$H_1^\perp$ and $f_{1T}^{\perp}$,
evaluated on the basis of a fit on experimental data in p$^\uparrow$p. 
Assuming the T-odd fragmentation function to
be responsible for the single spin asymmetry in pion production in
p$^\uparrow$p, we find the ratio $H_1^{\perp}/D_1$ to 
be in good agreement with the experimental results
from DELPHI data on $Z \, \to \, 2$-jet decay. 
We use our estimates to make predictions for ep$^\uparrow$.
\end{abstract}

  
\section{Introduction}

High energy scattering processes, e.g. unpolarized and  polarized 
deep inelastic scattering, provide an efficient tool to investigate the 
internal structure of nucleons. 
Particularly interesting is the study of the role 
that elementary constituents play in accounting for the total spin of 
the proton:   
 a joint theoretical and experimental effort is required to 
gain  complete knowledge and understanding of quark and gluon contributions 
to the spin structure of hadrons, in high energy processes.

At leading order in $1/Q$, the cross section for a hard process 
$a+b \to c+d$ is 
given by the convolution of a `hard part', which describes the scattering 
among elementary 
constituents and can be calculated perturbatively in the framework of QCD, 
and a `soft' part, that accounts for the processes in which either 
quarks are produced from the initial hadrons or final hadrons are produced 
from quarks resulting from the hard elementary scattering. 
Distribution functions belong to the first class of soft parts whereas 
fragmentation functions belong to the second class.

The most well-known distribution function, which we will indicate by 
$f^a_1(x)$, is the number density of quarks 
with flavour $a$ carrying a momentum fraction $x$ in an unpolarized proton; 
analogously, the fragmentation function $D_1^a(z_h)$ gives the 
density number of hadrons $h$ with momentum fraction $z_h$, resulting from the
fragmentation of a quark of flavour $a$.
When we consider polarized processes the number of distribution and 
fragmentation functions increases considerably. More specifically, we 
take into 
account the possibility of either extracting unpolarized 
quarks from polarized hadrons, or creating unpolarized or 
spinless hadrons from polarized quarks.

To distinguish among the various processes, we employ the following 
conventions (first introduced in Ref.~\cite{jj92} and later generalized in 
Ref.~\cite{tm9596}):
\begin{itemize}
\item
$f$ and $D$ apply to non polarized quarks in the proton or in the hadron 
respectively;
\item
$g$ and $G$ apply to longitudinally polarized quarks;
\item
$h$ and $H$ apply to transversely polarized quarks;
\item
the subscripts $L$ and $T$ refer to the 
longitudinal and transverse polarization of the target or produced 
(spin 1/2) hadron;
\item
A subscript $1$ indicates ``leading order'' 
(we will not deal with subleading functions here).
\item
In several polarized distribution and fragmentation functions, the 
intrinsic transverse momenta of quarks play an important role. In that
case a superscript ${\perp}$ is used to signal such a dependence on
$\bm k_\st$, while a superscript $(1)$ indicates that a function is integrated
over $\bm k_\st$, after a weighting with $\bm k_\st^2/2M^2$, see for example 
Eqs. (\ref{wf1Tperp},\ref{wH1Tperp}).
\end{itemize}

Figs~1 and~2 give a pictorial representation of these functions, and  
illustrate how the principles stated above are applied. 
The distribution function $f_1^a$ is the probability of finding an 
unpolarized quark $a$ into an unpolarized proton; this is a very well known object, usually determined by fits on unpolarized DIS experimental data.    
The distribution functions $g_{1L}^a$ and $g_{1T}^a$ are proportional to the 
probability of finding a quark with longitudinal polarization either in a 
longitudinally or in a transversely polarized proton, whereas  
$h_{1T}^a$ is proportional to the probability to find a transversely 
polarized quark $a$ in a transversely polarized hadron. 
In a completely analogous way, 
the fragmentation function $D_1^a$ is the probability of an 
unpolarized quark $a$ to fragment into an unpolarized hadron, whereas 
$G_{1L}^a$, $G_{1T}^a$, $H_{1T}^a$ take into account the probability 
of either longitudinally or transversely polarized quarks fragmenting into 
longitudinally or transversely polarized hadrons respectively.
In addition, we have distribution and fragmentation functions which are 
directly proportional to the intrinsic transverse momentum of the quarks 
inside the hadron; their contribution would then be zero in the approximation 
of zero intrinsic momentum. As shown in Figs.~1 and~2, $h_{1L}^{\perp a}$ 
and $h_{1T}^{\perp a}$ give the probability of a transversely polarized 
quark $a$  to be found in a longitudinaly or transversely polarized proton. 
Similarly, for fragmentation functions, we have $H_{1L}^{\perp}$ and 
$H_{1T}^{\perp}$. 

The distribution functions $f_{1T}^{\perp a}(x)$ and 
$h_{1}^{\perp a}(x)$, and the analogous fragmentation functions 
$D_{1T}^{\perp}(z)$ and $H_1^{\perp}(z)$ 
are particularly ``delicate'' and controversial objects. In fact, as it was 
extensively discussed in Ref. \cite{bm98}, those are T-odd functions 
(i.e. they are not constrained by time reversal invariance). 
This non-applicability of  time reversal 
symmetry is straightforwardly understood in the case of 
fragmentation functions, since the produced hadron can interact with the 
remnants of the fragmenting quark \cite{jj93}. 
Thus, a non-zero $H_1^{\perp}(z)$ allows 
for processes in which transversely polarized quarks fragment into 
unpolarized hadron (see picture in Fig.~2). 
Notice that,  as it was pointed out in Refs.~\cite{est98} and \cite{jbm98} 
a more accurate knowledge of this functions 
would give a unique chance to do spin physics with unpolarized or spin zero  
hadrons. 

In the case of the distribution functions,  
the non-application of  time reversal symmetry  
can still be accepted, due to soft initial state 
interactions~\cite{abm95} 
(it is, in fact, reasonable to believe that, in processes in which two 
hadrons are in the initial state, debris from 
the ``distribution'' process may soft-interact, mutually and with the 
quark which will be involved in the hard scattering) 
or, possibly, as a consequence of chiral symmetry breaking, 
as suggested in Ref.~\cite{adm96}. Furthermore, they can also arise  
effectively from higher order processes where soft gluons may produce 
so-called gluonic poles~\cite{gluonicpoles} . 
All in all, the $f_{1T}^{\perp a}(x)$ function being non-zero allows for 
processes in which unpolarized quarks are produced from a polarized proton
(see picture in Fig.~1).
 
But what about the real world? Are these effects really detectable in 
experiments? And what is the size of the effects generated by them? 
Investigating the $f_{1T}^{\perp a}(x)$ and $H_1^{\perp a}(z)$ is what this 
paper is about. 

For our estimates, we will benefit from two essential inputs: 
first of all we will use the parametrizations presented in Ref.~\cite{abm95} 
(or better those given in  the revamped version of Ref.~\cite{am98})   
and in Ref.~\cite{abm98}. In the first references, Anselmino {\it et al.} find 
an explicit parametrization for the $f_{1T}^{\perp a}(x)$ by fitting the 
data on single spin asymmetry in $p^{\uparrow}\,p \,\to \, \pi \,X$ 
from FNAL E704 experiment~\cite{adams}, assuming the presence of Sivers 
effect only~\cite{s90}, i.e. taking into account $k_{\perp}$ effects in the polarized 
proton initial state only. In  Ref.~\cite{abm98} the same authors present a 
parametrization of the 
$H_1^{\perp}(z)$ fragmentation function, based on a fit on the same 
experimental data, but taking into account only the Collins effect \cite{col}, 
thus assuming 
that the quark intrinsic transverse momentum has a relevant role in the final 
pion state kinematics only (see discussion in Ref.~\cite{abm98} for more 
details). \\
The second input we use is from Ref.~\cite{bm98}, where the 
authors explain how the T-odd fragmentation and distribution functions can 
be incorporated in their formalism and  suggest the use of some weighted 
integrals to get 
more information about them from measurements of specific angles ($\phi ^l _h$ 
and $\phi ^l _S$ in this particular case, which are the angle between the 
lepton scattering plane and the produced hadron plane, and the angle between 
the lepton scattering plane and the nucleon spin, respectively).

In Section~II we give a description of the formalism and notations and we 
analyze the relations between the correlators in the spin and 
helicity basis. We then discuss the connections between the distribution 
and fragmentation functions evaluated in Refs.~\cite{abm95,am98,abm98} and 
those presented in Ref.~\cite{bm98}. 
In Section~III the shape of some weighted integrals is shown as a function 
of $x$ and $z_h$ in tri-dimensional plots; they illustrate how time-reversal 
odd functions appear in some experimentally accessible observables. 
In the last Section we estimate the ratios $H_1^{\perp}/D_1$ 
and $f_{1T}^{\perp}/f_1$ and compare them with existing experimental 
data. In the concluding paragraph we will discuss the future perspectives 
and the experimental work that we recommend for a better and deeper 
understanding of these functions, and the interesting physics still hidden 
inside them.

\section{Quark correlation functions}

\subsection{Definitions for distribution functions}

%%%%%%%%%%%%%%%%%%%%%%%%%%%%%%%%%%%%%%%%%%%%%%%%%%%%%%%%%%%%%%%%%%%%%%%%%%
\begin{figure}[t]
\begin{center}
\mbox{~\epsfig{file=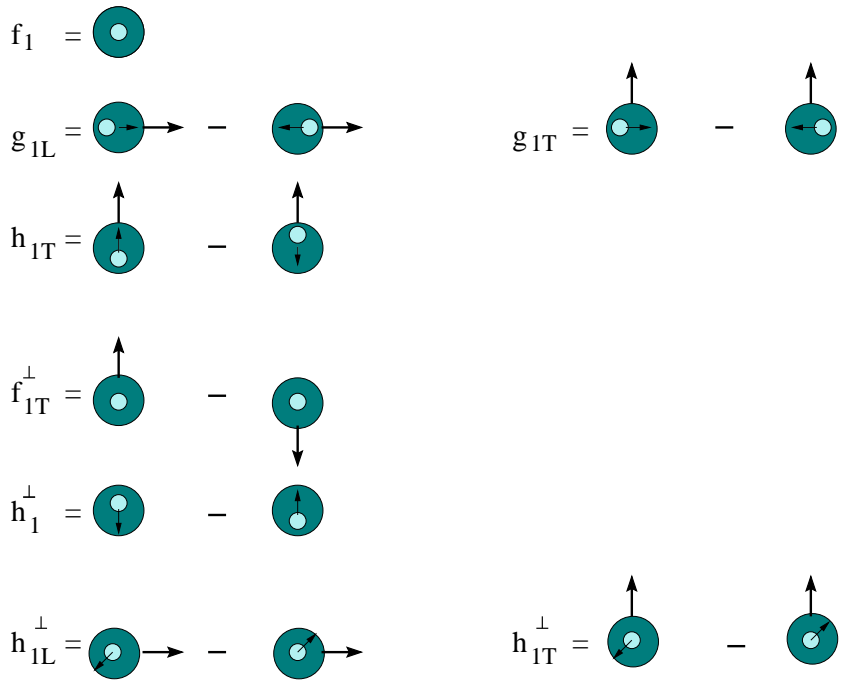,angle=0,width=11cm}}
\vspace{0.6cm}
\caption{\label{picture1}
Pictorial representation of the various kinds of distribution functions}
\end{center}
\end{figure}
%%%%%%%%%%%%%%%%%%%%%%%%%%%%%%%%%%%%%%%%%%%%%%%%%%%%%%%%%%%%%%%%%%%%%%%%%%%
%
%%%%%%%%%%%%%%%%%%%%%%%%%%%%%%%%%%%%%%%%%%%%%%%%%%%%%%%%%%%%%%%%%%%%%%%%%%
\begin{figure}[t]
\begin{center}
\mbox{~\epsfig{file=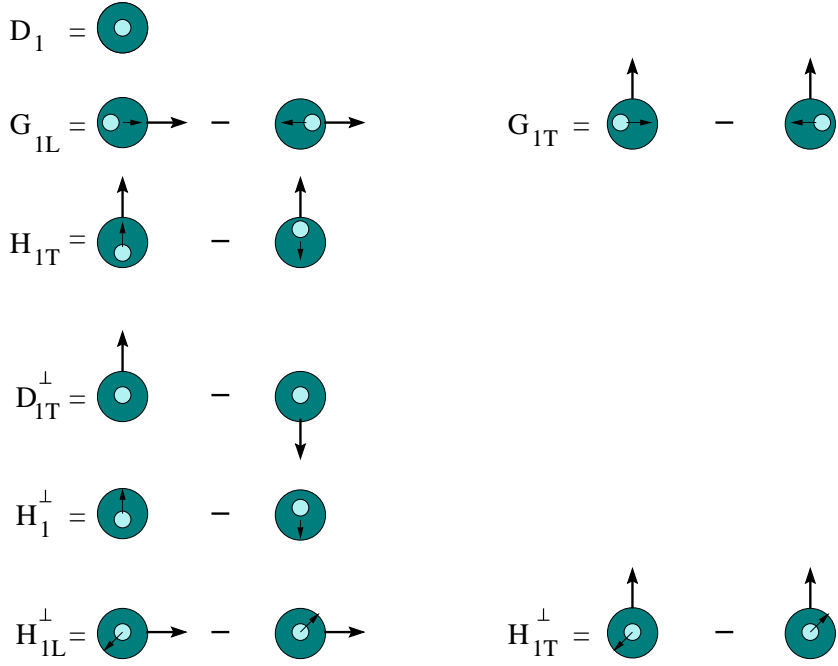,angle=0,width=11cm}}
\vspace{0.6cm}
\caption{\label{picture2}
Pictorial representation of the various kinds of fragmentation functions}
\end{center}
\end{figure}
%%%%%%%%%%%%%%%%%%%%%%%%%%%%%%%%%%%%%%%%%%%%%%%%%%%%%%%%%%%%%%%%%%%%%%%%%%%

The quark distribution functions alluded to in the introduction appear in
the parametrization of the lightfront correlation function~\cite{correlation}
\be
\Phi_{ij}(x,\bm k_\st; P,S) 
= \frac{1}{2}\,\left. \int \frac{d\xi^-d^2\bm \xi_\st}{(2\pi)^3}
\ e^{ik\cdot \xi} \,\langle P,S \vert \overline \psi_j (0)
\psi_i(\xi) \vert P,S \rangle \right|_{\xi^+ = 0},
\label{correlator}
\ee
which depends on the lightcone fraction of the quark momentum, $x = k^+/P^+$
and the transverse momentum component $\bm k_\st$. For this purpose we
use lightlike vectors $n_\pm$, satisfying $n_+\cdot n_- = 1$ and 
defining the lightcone coordinates $a^\pm$ = $a\cdot n_\mp$. The 
lightlike vectors are defined by the hadron momentum, $P$ $\equiv$
$P^+n_+ + (M^2/2P^+)n_-$, where $n_-$ is defined via another vector in the 
hard scattering process, e.g. the momentum transfer $q$ in inclusive deep 
inelastic scattering or the `other' hadron momentum in pp scattering. 
The definitions of $x$ and $\bm k_\st$ are contained in $k$ $\equiv$ 
$x\,P^+n_+ + k^-n_- + k_\st$.

Using Lorentz invariance, hermiticity, and parity invariance one finds that
the Dirac structure relevant in a calculation up to
leading order in $1/Q$ is given by \cite{bm98}
\bea
\Phi(x,\bm k_\st; P,S) &=& \frac{1}{4}\Biggl\{
f_1\,\slash \slash n_+
+ f_{1T}^\perp\, \frac{\epsilon_{\mu \nu \rho \sigma}
\gamma^\mu n_+^\nu k_\st^\rho S_\st^\sigma}{M}
+g_{1s}\, \gamma_5\slash n_+
\nonumber \\ & & \qquad
+h_{1T}\,i\sigma_{\mu\nu}\gamma_5 n_+^\mu S_\st^\nu
+h_{1s}^\perp\,\frac{i\sigma_{\mu\nu}\gamma_5
n_+^\mu k_\st^\nu}{M}
+ h_1^\perp\,\frac{\sigma_{\mu \nu} k_\st^\mu n_+^\nu}{M}
\Biggr\},
\label{param}
\eea
with arguments $f_1$ = $f_1(x,\bm k_\st^2)$ etc. 
Note that the factor $1/2$ in Eq.~(\ref{correlator}) and the parametrization
of Eq.~(\ref{param}) are chosen 
to get the proper normalization of the distribution functions, 
$\int dx\ d^2 \bm k_\st\ f_1^a(x,\bm k_\st)=n_a$,
from the relation
$\langle P,S\vert \overline \psi(0)\gamma^+\psi(0)\vert P,S\rangle$
= $2P^+\,n_a$.
The quantity $g_{1s}$ (and similarly $h_{1s}^\perp$) is shorthand for
\be
g_{1s}(x,\bm k_\st) =
\lambda\,g_{1L}(x,\bm k_\st^2) + \frac{\bm k_\st\cdot \bm S_\st}{M}
\,g_{1T}(x,\bm k_\st^2),
\ee
with $M$ the mass, $\lambda = M\,S^+/P^+$ the lightcone helicity,
and $\bm S_\st$ the transverse spin of the target hadron. In fact, we 
have $S$ $\equiv$ $\lambda\,(P^+/M)\,n_+ - \lambda\,(M/2P^+)\,n_- + S_\st$ 
and thus in the the restframe $S = (0,\bm S_\st,\lambda)$.
The lightcone helicity, thus, is a convenient quantity which in the target rest
frame is just the third component of the spin vector, while in the
infinite momentum frame ($P^+ \rightarrow \infty$) it is proportional 
to the standard helicity.

\subsection{Correlators in helicity basis}

In order to compare with other results we give the 
link with the helicity formalism, used in Refs.~\cite{abm95,abm98}.
This is achieved by transforming the $\Phi_{ij}$ matrix elements to the 
helicity basis with the help of the density matrix $\rho$, in the target rest 
frame given by
\be
\rho _{\Lambda \Lambda ^{\prime}} = \frac{1}{2} \;
( \delta _{\Lambda\Lambda^\prime}
+ \bm S \cdot (\bm \sigma)_{\Lambda \Lambda^\prime}) \; ,
\ee
where $\Lambda$, $\Lambda ^{\prime}$ are the helicity indices of the proton
and $S$ the spin vector used in the above expression.
In fact the parametrization using the spin vector $S$ is defined
as 
\be
\Phi_{ij} (x,k_\st;P,S) = \sum_{\Lambda \Lambda ^{\prime}}
\rho _{\Lambda \Lambda ^{\prime}} (S)
\; \Phi _{\Lambda i;\, \Lambda ^{\prime} j} (x,k_\st;P).
\ee
Using the restframe result $S = (0,\bm S_\st,\lambda)$ one obtains
\bea
\Phi_{ij}(x,\bm k_\st, P,S) & = &
\frac{1}{2}\Bigl(\Phi_{+i;\,+j} + \Phi_{-i;\,-j}\Bigr)
+\frac{1}{2}\,\lambda\Bigl(\Phi_{+i;\,+j} - \Phi_{-i;\,-j}\Bigr)
\nonumber \\
&&\mbox{}+\frac{1}{2}\,S_\st^1\Bigl(\Phi_{+i;\,-j} +\Phi_{-i;\,+j}\Bigr)
-\frac{i}{2}\,S_\st^2 \Bigl(\Phi_{+i;\,-j} - \Phi_{-i;\,+j}\Bigr) ,
\eea
where one could have used transverse spin differences instead of the 
off-diagonal helicity matrix elements. One immediately sees that 
\bea
&&\Bigl(\Phi_{+i;\,-j} +\Phi_{-i;\,+j}\Bigr)
= \Bigl(\Phi_{\uparrow_xi;\,\uparrow_xj} 
- \Phi_{\downarrow_xi;\,\downarrow_xj}\Bigr),
\\
&&-i\,\Bigl(\Phi_{+i;\,-j} -\Phi_{-i;\,+j}\Bigr)
= \Bigl(\Phi_{\uparrow_yi;\,\uparrow_yj} 
- \Phi_{\downarrow_yi;\,\downarrow_yj}\Bigr).
\eea

The Dirac structure of the above matrix elements can be translated into quark 
chiralities (for massless quarks, helicities) or transverse spin,
by using the appropriate Dirac projection operators, $P_{R/L} = (1\pm 
\gamma_5)/2$ or $P_{\uparrow_i/\downarrow_i}$ $= (1\pm 
\gamma^i\gamma_5)/2$ respectively, in combination with the projector 
onto the so-called good components, $P_+=\gamma ^- \gamma^+ /2$. 
Explicitly, for $\Phi_{ij}$ we can relate specific projections
$\mbox{Tr}\left(\Phi\,\Gamma\right)$
to transverse spin matrix elements or off-diagonal
quark chirality matrix elements $\Phi_{\lambda\lambda^\prime}$
(see Ref.~\cite{jj92}),
%\bea
%&&\Phi^{[\gamma^+]}(x,\bm k_\st;P,S) =
%\mbox{Tr}\,\Bigl( \Phi\,\gamma^+\Bigr)
%\equiv \Phi_{++} + \Phi_{--},
%\\
%&&\Phi^{[\gamma^+ \gamma_5]}(x,\bm k_\st;P,S) =
%\mbox{Tr}\,\Bigl( \Phi\,\gamma^+\gamma_5\Bigr)
%\equiv \Phi_{++} - \Phi_{--},
%\\
%&&\Phi^{[ i \sigma^{1+} \gamma_5]}(x,\bm k_\st;P,S) =
%\mbox{Tr}\,\Bigl( \Phi\,i \sigma^{1+} \gamma_5\Bigr)
%\equiv \Phi_{\uparrow_x\uparrow_x} - \Phi_{\downarrow_x\downarrow_x}
%= \Phi_{+-} + \Phi_{-+},
%\\
%&&\Phi^{[ i \sigma^{2+} \gamma_5]}(x,\bm k_\st;P,S) =
%\mbox{Tr}\,\Bigl( \Phi\,i \sigma^{2+} \gamma_5\Bigr)
%\equiv \Phi_{\uparrow_y\uparrow_y} - \Phi_{\downarrow_y\downarrow_y}
%= -i\Bigl(\Phi_{+-} + \Phi_{-+}\Bigr).
%\eea
%or
\bea
&&\Phi^{[\gamma^+]}(x,\bm k_\st;P,S) =
\mbox{Tr}\,\Bigl( \Phi\,\gamma^+\Bigr)
\equiv \Phi_{RR} + \Phi_{LL},
\\
&&\Phi^{[\gamma^+ \gamma_5]}(x,\bm k_\st;P,S) =
\mbox{Tr}\,\Bigl( \Phi\,\gamma^+\gamma_5\Bigr)
\equiv \Phi_{RR} - \Phi_{LL},
\\
&&\Phi^{[ i \sigma^{1+} \gamma_5]}(x,\bm k_\st;P,S) =
\mbox{Tr}\,\Bigl( \Phi\,i \sigma^{1+} \gamma_5\Bigr)
\equiv \Phi_{\uparrow_x\uparrow_x} - \Phi_{\downarrow_x\downarrow_x}
= \Phi_{RL} + \Phi_{LR},
\\
&&\Phi^{[ i \sigma^{2+} \gamma_5]}(x,\bm k_\st;P,S) =
\mbox{Tr}\,\Bigl( \Phi\,i \sigma^{2+} \gamma_5\Bigr)
\equiv \Phi_{\uparrow_y\uparrow_y} - \Phi_{\downarrow_y\downarrow_y}
= -i\Bigl(\Phi_{RL} - \Phi_{LR}\Bigr).
\eea
The Dirac projections applied to the parametrization in Eq.~(\ref{param}) 
gives
\bea
&&\Phi^{[\gamma^+]}(x,\bm k_\st;P,S)
%\mbox{Tr}\,\Bigl( \Phi\,\gamma^+\Bigr)
= f_1(x ,\bm k_\st)
- \frac{\epsilon_\st^{ij}k_{\st i} S_{\st j}}{M}\,f_{1T}^\perp(x,\bm k_\st),
\\
&&\Phi^{[\gamma^+ \gamma_5]}(x,\bm k_\st;P,S)
%\mbox{Tr}\,\Bigl( \Phi\,\gamma^+\gamma_5\Bigr)
= \lambda\,g_{1L}(x ,\bm k_\st)
+ g_{1T}(x ,\bm k_\st)\,\frac{(\bm k_\st\cdot\bm S_\st)}{M} ,
\\
&&\Phi^{[ i \sigma^{i+} \gamma_5]}(x,\bm k_\st;P,S) 
%\mbox{Tr}\,\Bigl( \Phi\,i \sigma^{i+} \gamma_5\Bigr)
= S_\st^i\,h_{1}(x ,\bm k_\st)
+ \frac{\lambda k_\st^i}{M} \,h_{1L}^\perp(x ,\bm k_\st)
\nonumber \\ && \hspace{4cm} \mbox{}
- \frac{\left(k_\st^i k_\st^j + \frac{1}{2}\bm k_\st^2g_\st^{ij}\right)
S_{\st j}}{M^2} \,h_{1T}^\perp(x ,\bm k_\st)
- \frac{\epsilon_\st^{ij} k_{\st j}}{M}\,h_1^\perp(x,\bm k_\st).
\eea
In the final equation for the transverse spin 
distributions the combination $h_1$ is, in fact, $h_1$ = $h_{1T} 
+ (\bm k_\st^2/2M^2)\,h_{1T}^\perp$ because it is this combination
which survives after integration over $\bm k_\st$. 
The expressions provide the appropriate interpretation 
of the distribution functions as illustrated in Fig.~\ref{picture1}

Combining the nucleon helicities instead of the parametrization with the spin
vector and quark chiralities instead of the Dirac structure, one can
immediately transform the functions appearing in the projections above 
into matrix elements 
$\Phi_{\Lambda\lambda;\Lambda^\prime\lambda^\prime}(x,k_\st;P)$,
which we give here for completeness,
\bea
&&
f_1 \ = \ 
\frac{1}{2}\Biggl(
\Phi_{+R;+R} + \Phi_{+L;+L} + \Phi_{-R;-R} + \Phi_{-L;-L} 
\Biggr),
\\ &&
\frac{\vert \bm k_\st\vert \,\sin \phi}{M}\,f_{1T}^\perp \ = \ 
\frac{1}{2}\Biggl(
\Phi_{+R;-R} + \Phi_{+L;-L} + \Phi_{-R;+R} + \Phi_{-L;+L} 
\Biggr),
\\ &&
-\frac{\vert \bm k_\st\vert \,\cos \phi}{M}\,f_{1T}^\perp \ = \ 
-\frac{i}{2}\Biggl(
\Phi_{+R;-R} + \Phi_{+L;-L} - \Phi_{-R;+R} - \Phi_{-L;+L} 
\Biggr),
\\ &&
g_{1L} \ = \ 
\frac{1}{2}\Biggl(
\Phi_{+R;+R} - \Phi_{+L;+L} - \Phi_{-R;-R} + \Phi_{-L;-L} 
\Biggr),
\\ &&
\frac{\vert \bm k_\st\vert \,\cos \phi}{M}\,g_{1T} \ = \ 
\frac{1}{2}\Biggl(
\Phi_{+R;-R} - \Phi_{+L;-L} + \Phi_{-R;+R} - \Phi_{-L;+L} 
\Biggr),
\\ &&
\frac{\vert \bm k_\st\vert \,\sin \phi}{M}\,g_{1T} \ = \ 
-\frac{i}{2}\Biggl(
\Phi_{+R;-R} - \Phi_{+L;-L} - \Phi_{-R;+R} + \Phi_{-L;+L} 
\Biggr),
\\ &&
\frac{\vert \bm k_\st\vert \,\sin \phi}{M}\,h_1^\perp \ = \ 
\frac{1}{2}\Biggl(
\Phi_{+R;+L} + \Phi_{+L;+R} + \Phi_{-R;-L} + \Phi_{-L;-R} 
\Biggr),
\\ &&
-\frac{\vert \bm k_\st\vert \,\cos \phi}{M}\,h_1^\perp \ = \ 
-\frac{i}{2}\Biggl(
\Phi_{+R;+L} - \Phi_{+L;+R} + \Phi_{-R;-L} - \Phi_{-L;-R} 
\Biggr),
\\ &&
\frac{\vert \bm k_\st\vert\,\cos\phi}{M}\,h_{1L}^\perp \ = \ 
\frac{1}{2}\Biggl(
\Phi_{+R;+L} + \Phi_{+L;+R} - \Phi_{-R;-L} - \Phi_{-L;-R} 
\Biggr),
\\ &&
\frac{\vert \bm k_\st\vert\,\sin\phi}{M}\,h_{1L}^\perp \ = \ 
-\frac{i}{2}\Biggl(
\Phi_{+R;+L} - \Phi_{+L;+R} - \Phi_{-R;-L} + \Phi_{-L;-R} 
\Biggr),
\\ &&
h_1 + \frac{\vert\bm k_\st\vert^2 \,\cos 2\phi}{2M^2}\,h_{1T}^\perp \ = \ 
\Biggl(
\Phi_{+L;-R} + \Phi_{-R;+L} 
\Biggr),
\\ &&
\frac{\vert\bm k_\st\vert^2 \,\sin 2\phi}{2M^2}\,h_{1T}^\perp \ = \ 
-i\Biggl(
\Phi_{+R;-L} - \Phi_{-L;+R} 
\Biggr),
\eea
where $\phi$ is the azimuthal angle of the quark transverse momentum.

\subsection{Explicit evaluation of time-reversal odd distribution functions}

%%%%%%%%%%%%%%%%%%%%%%%%%%%%%%%%%%%%%%%%%%%%%%%%%%%%%%%%%%%%%%%%%%%%%%%%%%
\begin{figure}[t]
\begin{center}
\mbox{~\epsfig{file=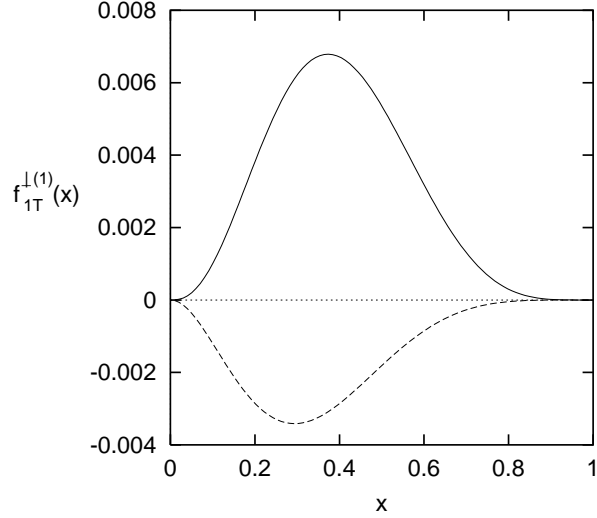,angle=-90,width=8cm}}
\vspace{0.6cm}
\caption{\label{f1T1}
The $(\bm k_\st^2/2M^2)$-moment of the T-odd distribution function,
$f_{1T}^{\perp \, (1)\,  u}(x)$,
solid line, and $f_{1T}^{\perp \, (1) \, d}(x)$, dashed line, 
evaluated from Eq. (\protect \ref{f(1)}). }
\end{center}
\end{figure}
%%%%%%%%%%%%%%%%%%%%%%%%%%%%%%%%%%%%%%%%%%%%%%%%%%%%%%%%%%%%%%%%%%%%%%%%%%%
%

{}From these expressions, one can easily see that the term proportional to
$f_{1T}^{\perp}$ in the $\Phi _{ij}^{[\gamma ^{+}]}$ projection can be
identified with the function $\Delta^Nf_{q/\uparrow}$ = $2\,I_{+-}$ 
defined in Ref.~\cite{abm95}.
To be more precise, one finds
\be
% 2/,I_{+-}(x,\bm k_\st) = 
\Delta^Nf_{q/\uparrow}(x,\bm k_\st) 
= 2\,\frac{\vert \bm k_\st\vert\,\sin\phi}{M} \; 
f_{1T}^{\perp}(x,\bm k_\st).
\label{I+-}
\ee
In later applications it will turn out to be useful to consider the
$(\bm k_\st^2/2M^2)$ weighted function
\be
f_{1T}^{\perp (1)}(x) = \int d^2k_\st\ \frac{\vert \bm 
k_\st\vert^2}{2M^2}\,f_{1T}^\perp (x,\bm k_\st)\;,
\label{wf1Tperp}
\ee
for which we use the estimate
\be
f_{1T}^{\perp (1)}(x)  
%=\frac{\langle k_\st(x)\rangle}{2M}\,I_{+-}(x)
= \frac{\langle k_\st(x)\rangle}{4M}\,\Delta^Nf_{q/\uparrow}
(x)\;.
\label{f(1)}
\ee
Using the results from the
most recent analysis of the pion left-right asymmetry in 
p$^\uparrow$p $\rightarrow \pi$X in Ref.~\cite{am98},
\bea
&& \Delta ^N f_{u/\uparrow}(x) = 6.90\,x^{2.02}\,(1-x)^{4.06},\\
&& \Delta ^N f_{d/\uparrow}(x) = -2.34\,x^{1.44}\,(1-x)^{4.62},
\eea
and the results from, for example, Ref.~\cite{jrr89} for the average
transverse momentum,
\be
\frac{<k_\st(x)>}{M} = 0.47 \, x ^{0.68} \, (1-x)^{0.48} \; ,
\ee
we obtain for
$f_{1T}^{\perp (1)}$ the estimate
\bea
f_{1T}^{\perp (1)\,u}(x) &=& 0.81\;x^{2.70}\;(1-x)^{4.54}\; , \nonumber \\
f_{1T}^{\perp (1)\,d}(x) &=& -0.27\;x^{2.12}\;(1-x)^{5.10} .
\label{f1}
\eea
These estimates are shown in Fig.~\ref{f1T1}.

\subsection{Definitions and correlators for fragmentation functions}

For fragmentation functions one can proceed in an analogous way.
The quark fragmentation functions alluded to in the introduction appear in
the parametrization of the lightfront correlation function
\be
\Delta_{ij}(z,\bm k_\st,P_h) =  \frac{1}{2z}\left. \int 
\frac{d\xi^+d^2\bm \xi_\st}{(2\pi)^3}
\ e^{ik\cdot \xi} \,\langle 0 \vert\psi_i (\xi)\vert P_h\rangle
\,\langle P_h;X\vert
\psi_j(0) \vert 0 \rangle \right|_{\xi^- = 0}.
\ee
They depend on the lightcone fraction of the quark momentum, $z = P_h^-/k^-$
and the transverse momentum component $\bm k_\st$. The `dominant' 
direction is choosen to be the minus direction in this case. For the 
transverse directions we note that one has
\be
k_\st = k - \frac{P_h}{z} + (\ldots)\,n_+\;,
\ee
up to an (irrelevant) plus-component. This shows that we can 
interpret $k_\st$ as the quark transverse momentum in a frame where 
the produced hadron has no transverse component, while we can 
interpret $k_\st^\prime = -z\,k_\st$ as the transverse momentum 
of the produced hadron in a frame where the fragmenting quark has 
no transverse momentum.
 
Just as for the distribution functions,
the full Dirac structure relevant for fragmentations has been given in 
Ref.~\cite{bm98}. We limit ourselves to fragmention 
into spin 0 (or unpolarized) 
hadrons is given. Up to leading order in $1/Q$ the result is
\bea
\Delta(z,\bm k_\st,P_h) &=&
\frac{1}{2}\Biggl\{
D_1\,\slash \slash n_-
%+ D_{1T}^\perp\, \frac{\epsilon_{\mu \nu \rho \sigma}
%\gamma^\mu n_-^\nu k_\st^\rho S_{h\st}^\sigma}{M_h}
%+G_{1s}\, \gamma_5\slash n_-
%\nonumber \\ & & \qquad
%+H_{1T}\,i\sigma_{\mu\nu}\gamma_5 n_-^\mu S_{h\st}^\nu
%+H_{1s}^\perp\,\frac{i\sigma_{\mu\nu}\gamma_5
%n_-^\mu k_\st^\nu}{M_h}
+ H_1^\perp\,\frac{\sigma_{\mu \nu} k_\st^\mu n_-^\nu}{M_h}
\Biggr\},
\label{DDelta}
\eea
where $M_h$ is the mass of the produced hadron, and the arguments of $D_1$ 
and $H_1^{\perp}$ are $z$ and $\bm k_\st^{\prime 2}$.  
%The quantity $G_{1s}$(and similarly $H_{1s}^\perp$) is shorthand for
%\be
%G_{1s}(z,\bm k_\st^\prime) =
%\lambda_h\,G_{1L}(z,\bm k_\st^{\prime 2}) 
%+ \frac{\bm k_\st\cdot \bm S_{h\st}}{M_h}
%\,G_{1T}(x,\bm k_\st^{\prime 2}),
%\ee
%with $\lambda_h = M_h\,S_h^-/P_h^-$ the lightcone helicity,
%and $\bm S_{h\st}$ the transverse spin of the produced hadron. In fact, we 
%have $S_h$ $\equiv$ $\lambda_h\,(P_h^-/M_h)\,n_- 
%- \lambda_h\,(M_h/2P_h^-)\,n_+ + S_{h\st}$ 
%and thus in the the restframe $S_h = (0,\bm S_{h\st},\lambda_h)$.
%In order to compare with other results we give the link with
%the helicity formalism..
The normalization is fixed via the momentum sum rule 
$\sum_h \int dz\,d^2k_\st^\prime\ z\,D_1^{(h/q)}(z,\bm k_\st^\prime) = 1$.

For the interpretation in terms of quark chiralities one needs to
consider the Dirac projections
\bea
&&\frac{1}{2}\,\Delta^{[\gamma^-]}(z,\bm k_\st,P_h)
\equiv \frac{1}{2}\,\mbox{Tr}\,\Bigl( \Delta \,\gamma^-\Bigr)
= \frac{1}{2}\Bigl(\Delta_{RR} + \Delta_{LL} \Bigr)
\equiv {\cal N}_{h/q}(z,\bm k_\st^\prime),  
\\
&&\frac{1}{2}\,\Delta^{[\gamma^- \gamma_5]}(z,\bm k_\st)
\equiv 
\frac{1}{2}\,\mbox{Tr}\,\Bigl(\Delta \,\gamma^-\gamma_5\Bigr)
= \frac{1}{2}\Bigl(\Delta_{RR} - \Delta_{LL} \Bigr)
\equiv {\cal N}_{h/q}(z,\bm k_\st^\prime)\,\lambda_q(z,\bm k_\st^\prime),
\\
&&\frac{1}{2}\,\Delta^{[ i \sigma^{1-} \gamma_5]}(z,\bm k_\st,P_h) \equiv 
\frac{1}{2}\,\mbox{Tr}\,\Bigl(\Delta \,i\sigma^{1-}\gamma_5\Bigr)
= \frac{1}{2}\Bigl(\Delta_{\uparrow_x\uparrow_x} 
- \Delta_{\downarrow_x\downarrow_x} \Bigr)
\nonumber \\ &&\mbox{} \hspace{5.5cm}
= \frac{1}{2}\Bigl(\Delta_{RL} + \Delta_{LR} \Bigr)
\equiv {\cal N}_{h/q}(z,\bm k_\st^\prime)\,s_q^1(z,\bm k_\st^\prime),
\\
&&\frac{1}{2}\,\Delta^{[ i \sigma^{2-} \gamma_5]}(z,\bm k_\st,P_h) \equiv 
\frac{1}{2}\,\mbox{Tr}\,\Bigl(\Delta \,i\sigma^{2-}\gamma_5\Bigr)
= \frac{1}{2}\Bigl(\Delta_{\uparrow_y\uparrow_y} 
- \Delta_{\downarrow_y\downarrow_y} \Bigr)
\nonumber \\ &&\mbox{} \hspace{5.5cm}
= -\frac{i}{2}\Bigl(\Delta_{RL} - \Delta_{LR} \Bigr)
\equiv {\cal N}_{h/q}(z,\bm k_\st^\prime)\,s_q^2(z,\bm k_\st^\prime).
\eea
For the production of unpolarized hadrons, we obtain from Eq.~(\ref{DDelta})
\bea
&&\frac{1}{2}\,\Delta^{[\gamma^-]}(z,\bm k_\st,P_h)
\ = \ D_1(z,\bm k_\st^\prime),
\\
&&\frac{1}{2}\,\Delta^{[\gamma^- \gamma_5]}(z,\bm k_\st,P_h) \ = \ 0 ,
\\
&&\frac{1}{2}\,\Delta^{[ i \sigma^{i-} \gamma_5]}(z,\bm k_\st,P_h) \ = \  
\frac{\epsilon_\st^{ij} k_{\st j}}{M_h}\,H_1^\perp(z,\bm k_\st^\prime).
\eea
Explicitely,
\bea
&&
D_1 \ = \ \frac{1}{2}\,(\Delta _{RR}+\Delta _{LL}),
\\ &&
-\frac{\vert \bm k_\st\vert \,\sin \phi}{M_h}\,H_{1}^\perp \ = \ 
\frac{1}{2}\,(\Delta_{RL} + \Delta_{LR}),
\\ &&
\frac{\vert \bm k_\st\vert \,\cos \phi}{M_h}\,H_{1}^\perp \ = \ 
-\frac{i}{2}\,(\Delta_{RL} - \Delta_{LR}).
\eea

\subsection{Explicit evaluation of time-reversal odd fragmentation functions}

The fragmentation function $H_1^\perp$ describes the production of 
unpolarized hadrons, e.g. pseudoscalar mesons, from transversely polarized 
quarks. It is related to the
function $\Delta ^N D (z_h)$ in Ref.~\cite{abm98} which is used to 
describe the
left-right asymmetry in p$^\uparrow$p $\rightarrow \pi$X. The precise
equivalence is
\be
\Delta ^N D (z,\bm k_\st)\,d^2 \bm k_ \st = -2\,\frac{\vert 
\bm k_\st\vert\,\sin\phi}{M_h}
\,H_1^{\perp}(z,\bm k_\st^\prime)\, d^2 \bm k_ \st ^\prime \;,
\label{Delta}
\ee
where $\phi$ is the relative azimuthal angle of the
outgoing hadron momentum.
%
%%%%%%%%%%%%%%%%%%%%%%%%%%%%%%%%%%%%%%%%%%%%%%%%%%%%%%%%%%%%%%%%%%%%%%%%%%
\begin{figure}[t]
\begin{center}
\mbox{~\epsfig{file=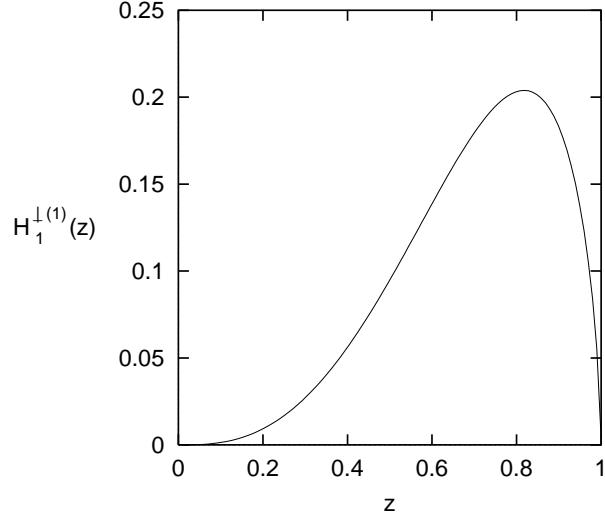,angle=-90,width=8cm}}
\vspace{0.6cm}
\caption{\label{fig4}
The T-odd fragmentation function first moment,
$H_{1}^{\perp \, (1)\,\mbox{\small fav}}(z)$, as it can
be evaluated from Eq.~(\protect \ref{H(1)}). }
\end{center}
\end{figure}
%%%%%%%%%%%%%%%%%%%%%%%%%%%%%%%%%%%%%%%%%%%%%%%%%%%%%%%%%%%%%%%%
%
In later applications we will use
\be
H_1^{\perp (1)}(z) = \int d^2k_\st^\prime
\ \frac{\vert \bm k_\st\vert^2}{2M_h^2}
\,H_1^\perp (z,\bm k_\st^\prime),
\label{wH1Tperp}
\ee
for which we use the estimate
\be
H_1^{\perp (1)}(z) = -\frac{\langle k_\st(z)\rangle}{4M_h}
\,\Delta ^N D(z).
\ee
We now make use of the results of Ref.~\cite{abm98},
\be
\Delta ^N D(z) = -0.13\,z^{2.60}\,(1-z)^{0.44},
\ee
and of a fit to the
LEP data~\cite{abreu},
\be
\frac{<k_\st(z)>}{M_{\mbox{\footnotesize ref}}} = 0.61 \, z^{0.27} \, (1-z)^{0.20} 
\; ,
\ee
where $M_{\mbox{ref}}=1$ GeV. 
Taking into account that the $H_1^{\perp (1)}(z)$ is scaled to the mass 
of the produced hadron, a pion in this specific case, we get
\be
H_{1}^{\perp (1)}(z) = 1.08\;z^{2.87}\;(1-z)^{0.64}\;.
\label{H(1)}
\ee
This is the result for the favored fragmentation functions for which we
have imposed isospin symmetry
\bea
&&H_{1}^{\perp\,\mbox{\small fav}}=
H_{1}^{\perp\ u\rightarrow \pi^+} =  H_{1}^{\perp\ \bar d\rightarrow \pi^+} 
= H_{1}^{\perp\ d\rightarrow \pi^-} =  H_{1}^{\perp\ \bar u\rightarrow \pi^-} 
= 2\, H_{1}^{\perp\ q\rightarrow \pi^0} \; ;\\
&&H_{1}^{\perp\,\mbox{\small non-fav}} =
H_{1}^{\perp\ d\rightarrow \pi^+} =  H_{1}^{\perp\ \bar u\rightarrow \pi^+} 
= H_{1}^{\perp\ u\rightarrow \pi^-} =  H_{1}^{\perp\ \bar d\rightarrow \pi^-} 
= 0 \; .
\eea
In Fig.~\ref{fig4} we show the function $H_{1}^{\perp\,(1)\,\mbox{\small fav}}
(z)$.
Notice that the T-odd
distribution function $f_{1T}^{\perp\,(1)}(x)$, reaches its maximum
for relatively small values of $x$, whereas the fragmentation function
$H_{1}^{\perp\,(1)}(z)$ has a maximum for a large value of $z$.

In order to make estimates for leptoproduction cross sections we 
need also an estimate for the polarized distribution functions $h_1(x)$.
As in Ref.~\cite{abm98} we assume
\bea
h_{1}^{u}(x) &=& P^{u/p^{\uparrow}}\,f_{1}^u(x) \; , \nonumber \\
h_{1}^{d}(x) &=& P^{d/p^{\uparrow}}\,f_{1}^d(x) \; .
\eea 
Here the polarization factors $P^{u/p^{\uparrow}}$ and $P^{d/p^{\uparrow}}$ 
are defined as
$P^{u/p^{\uparrow}} = P^{u^{\uparrow}/p^{\uparrow}} - 
P^{u^{\downarrow}/p^{\uparrow}} \;$ and 
$P^{d/p^{\uparrow}} = P^{d^{\uparrow}/p^{\uparrow}} - 
P^{d^{\downarrow}/p^{\uparrow}}$,
and they will be taken from SU(6) flavour symmetry estimates
\be
P^{u/p^{\uparrow}} = 2/3 \;\; , \;\;  P^{d/p^{\uparrow}} = -1/3\;.
\ee
%or from Ref.~\cite{abm98} fit results
%\be
%P^{u/p^{\uparrow}} = 2/3 \;\; , \;\;  P^{d/p^{\uparrow}} = -0.88; .
%\ee
The unpolarized distribution functions $f_{1}^u(x)$ and $D_1(z_h)$ are 
available in various styles and versions in the literature. We choose the 
MRSG~\cite{mrsg} set for $f_{1}^u(x)$ and the LO Binnewies {\it et al.} 
set~\cite{binn} for $D_1(z_h)$.

\section{Evaluation of weighted integrals}

We now have all the ingredients to calculate the weighted integrals proposed 
in Ref.~\cite{bm98}. Following the notations introduced therein,
we will focus our attention on three of such objects:
\begin{enumerate}
\item 
First of all we will consider, as a term of reference, the cross-section 
corresponding to a fully unpolarized DIS process, which is simply obtained 
by contracting the lepton tensor with the hadronic tensor (see Eq.~(16) 
in Ref.\cite{bm98}). Then we find the well known formula
\be
\langle 1 \rangle _{OOO} = \frac{ 4 \pi \alpha ^2 s}{Q^4} 
\left(1-y+\frac{y^2}{2} \right) \sum _{a,\overline a} e_a ^2 x 
f_1^a(x) D_1 ^a(z_h)\;,
\label{fD}
\ee
where $y=Q^2/sx$ and $x$ is the Bjorken variable.
Here we applied the definition of weighted integrals given in 
Ref.~\cite{bm98}
\be
\langle W \rangle _{ABC} = \int d\phi ^l d^2 \bm q _T \, W \, 
\frac{d\sigma _{ABC} ^{l \, H \to l\, h \, X}}{dx\, dy\, dz_h \, d\phi ^l 
\, d^2 \bm q _T} \; ,
\ee
and the subscripts $A$, $B$ and $C$ denote the polarization of lepton, 
target hadron and produced hadron, respectively.
Fig.~5 shows a tridimensional plot of the quantity 
$\sum _{a,\overline a} e_a ^2 \ x  \, 
f_1^a(x) \, D_1 ^a(z_h)$ as a function of $x$ and $z_h$. Notice that this 
function is practically zero for large values of $x$ and $z_h$, whereas it 
very rapidly increases as $x$ and $z_h$ become smaller.  

%%%%%%%%%%%%%%%%%%%%%%%%%%%%%%%%%%%%%%%%%%%%%%%%%%%%%%%%%%%%%%%%%%%%%%%%%%
\begin{figure}[t]
\begin{center}
\mbox{~\epsfig{file=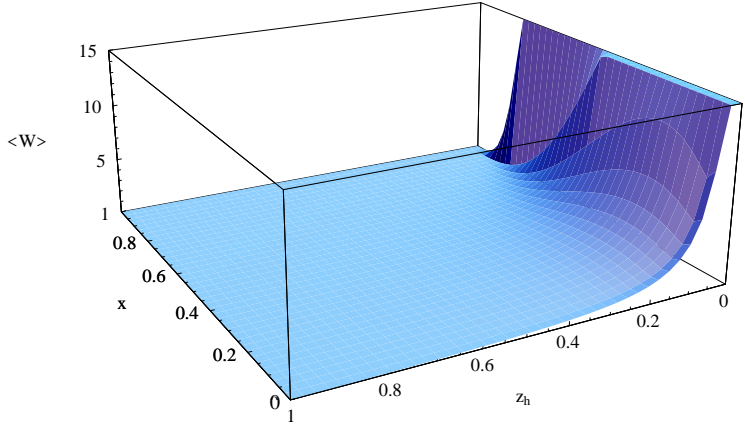,angle=0,width=10cm}}
\vspace{0.6cm}
\caption{A tri-dimensional view of the quantity 
$\sum _{a,\overline a} e_a ^2 \ x  \, f_1^a(x) \, D_1 ^a(z_h)$ 
as a function of $x$ and $z_h$. This represents the cross-section 
corresponding to a fully unpolarized DIS process (see Eq.~\protect \ref{fD}) 
leading to the production of a $\pi ^+$. 
Notice that only valence contributions are taken into account, 
for a consistent comparison with later plots. The cross-section becomes 
sizeable in the region in which both the variables $x$ and $z_h$ are 
relatively small.}
\end{center}
\end{figure}
%%%%%%%%%%%%%%%%%%%%%%%%%%%%%%%%%%%%%%%%%%%%%%%%%%%%%%%%%%%%%%%%%%%%%%%%%%%

\item
If we consider $W=(Q_T/M) \sin(\phi ^l _h - \phi ^l _S)$ in a scattering 
process with $ABC=OTO$, i.e. when an unpolarized beam hits a polarized proton 
target, we can single out a quantity which is directly proportional 
to our T-odd distribution function, or more precisely to its first moment, 
as given in Eq.~(\ref{f(1)}) (see also Table II, last line, 
in Ref.~\cite{bm98})
\be
\left< \frac{Q_T}{M}\;\sin(\phi^l_h - \phi^l_S) \right>_{OTO} = 
\frac{ 4 \pi \alpha ^2 s}{Q^4} \, (1-y)\,
\sum _{a,\overline a} e_a ^2 \ x \, f_{1T}^{\perp (1) a}(x) \, D_1 ^a(z_h)\;.
\label{f1D}
\ee
A tri-dimensional plot of the quantity $\sum _{a,\overline a} 
e_a ^2 \ x \, f_{1T}^{\perp (1) a}(x) \, D_1 ^a(z_h)$ is shown in Fig.~6. 
By comparing this weighted integral to the fully unpolarized cross-section, 
shown in Fig.~5, we see that this time the shape of the surface as a function 
of $x$ and 
$z_h$ has changed, since it becomes sizeable for very small values 
of $z_h$ and intermediate values of $x$. Notice also that the overall size of 
the function is considerably suppressed (by roughly two orders of magnitude) 
by the $<k_T>^2$ factor. Therefore, it is clear that the effects due to 
the presence of the T-odd  distribution function $f_{1T}^{\perp}(x)$ are 
small, but a suitably designed experiment may put limits on their existence, 
or might establish their mere existence. This would be a crucial test for 
the presence of T-odd distribution functions and provide a deeper 
understanding of these phenomena.

%%%%%%%%%%%%%%%%%%%%%%%%%%%%%%%%%%%%%%%%%%%%%%%%%%%%%%%%%%%%%%%%%%%%%%%%%%
\begin{figure}[t]
\begin{center}
\mbox{~\epsfig{file=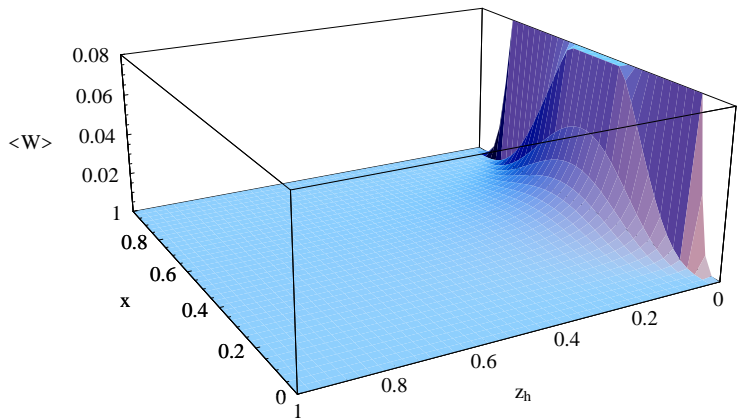,angle=0,width=10cm}}
\vspace{0.6cm}
\caption{A tri-dimensional view of the quantity 
$\sum _{a,\overline a} e_a ^2 \ x \, f_{1T}^{\perp (1) a}(x) \, D_1 ^a(z_h)$,
directly proportional to the T-odd distribution function $f_{1T}^{\perp}(x)$, 
see  Eq.~\protect (\ref{f1D}), for $OTO$ scattering with production of 
$\pi ^+$. 
Only valence contributions are taken into account. Here the function becomes 
sizeable for small values of $z_h$ but intermediate values of $x$. Notice that 
the overall size of the surface is considerably reduced by the action 
of the $<k_\st>^2$ factor.}
\end{center}
\end{figure}
%%%%%%%%%%%%%%%%%%%%%%%%%%%%%%%%%%%%%%%%%%%%%%%%%%%%%%%%%%%%%%%%%%%%%%%%%%%

\item 
Finally, if we choose the weight 
$W=(Q_T/M)\;\sin(\phi^l_h + \phi^l_S)$, we obtain an 
object which is directly proportional to the T-odd fragmentation function
$H_1^{\perp (1)}$ (see Table II, second line, in Ref.~\cite{bm98})
\be
\left< \frac{Q_T}{M}\;\sin(\phi ^l _h + \phi ^l _S) \right>_{OTO} =
\frac{ 4 \pi \alpha ^2 s}{Q^4} \, (1-y)\,
\sum _{a,\overline a} e_a ^2 \ x \, h_{1}^{a}(x)  \, H_1 ^{\perp (1) a} (z_h) 
\;.
\label{hH1}
\ee
As it clearly appears from the plot in Fig.~7, this time the 
shape of the quantity 
$\sum _{a,\overline a} e_a ^2 x \, h_{1}^{a}(x)  \, H_1 ^{\perp (1) a} (z_h)$ 
as a function of $x$ and $z_h$ is completely 
different from the previous two. It reaches its maximum 
for relatively small values of $x$ and for large values of $z_h$ and its 
overall size is at least a factor two bigger than the previous one. This 
means that a measure to reveal the effects of a non zero T-odd fragmentation 
function could easily be made at large values of $z_h$, where it is relatively 
easier to achieve  larger statistics.

%%%%%%%%%%%%%%%%%%%%%%%%%%%%%%%%%%%%%%%%%%%%%%%%%%%%%%%%%%%%%%%%%%%%%%%%%%
\begin{figure}[t]
\begin{center}
\mbox{~\epsfig{file=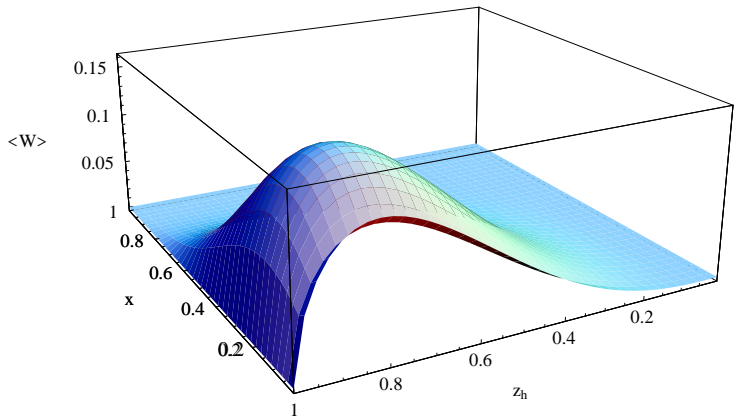,angle=0,width=10cm}}
\vspace{0.6cm}
\caption{A tri-dimensional view of $\sum _{a,\overline a} e_a ^2 x \, 
h_{1}^{a}(x)  \, H_1 ^{\perp (1) a} (z_h)$, directly 
proportional to the T-odd fragmentation function $H_{1}^{\perp}(z_h)$, see  
Eq.~\protect (\ref{hH1}), for $OTO$ scattering with production of $\pi ^+$. 
Once again, only valence contributions are taken into account. 
As opposed to the previous case, here the function reaches its maximum for  
considerbly large values of $z_h$.}
\end{center}
\end{figure}
%%%%%%%%%%%%%%%%%%%%%%%%%%%%%%%%%%%%%%%%%%%%%%%%%%%%%%%%%%%%%%%%%%%%%%%%%%%

\end{enumerate}

\vspace{1cm}

\small
\begin{center}
{\bf IV. EVALUATION OF THE RATIOS  
${\bm H_1 ^{\perp} / \bm D_1}$ \bf AND 
${\bm f_{1T} ^{\perp} / \bm f_1 }$}
\end{center}

\vspace{0.3cm}

\normalsize
We now focus our attention on the evaluation of the ratio 
$H_1 ^{\perp a}  /  D_1 ^a $, which will then be compared to the 
experimental results from DELPHI data on $Z \; \to \; 2$ jet decay, presented 
by Efremov {\it et al.} in Ref.~\cite{est98}. 
Once again we take $D_1(z_h)$ from the LO fragmentation function sets by 
Binnewies {\it et al.} \cite{binn} and $H_1^\perp(z_h)$ from Eq.~(\ref{H(1)}). To calculate 
this ratio, we have to fix the flavour, $a$, of the quark;  
so we start by considering, for instance, $\pi ^+$ production, in which $u$ 
is valence, and fix the flavour to be $u$ in our evaluation. 
As we can see, the integration over $z$ 
presents some technical problems, because the $D_1(z_h)$ fragmentation 
function diverges at small values of $z_h$. Then we will perform a cut at 
$z_h(min)=0.1$ (typical cuts in HERMES and COMPASS experiments)  
to get a finite result. Under these assumptions we have
\be
%\frac{H_1^{\perp\,\mbox{\small fav}}}{D_1^{u / \pi ^+}} = 
\left| \frac{\int_{0.1}^1 \, dz_h\ H_1^{\perp \mbox{\small fav}}(z_h)}
{\int_{0.1}^1 \, dz_h\ D_1^{u / \pi ^+} (z_h)} \right| = 0.076\;.
\ee
This means that our evaluation of the ratio $H_1 ^{\perp a}  /  D_1 ^a $ 
gives  a value of about $8\%$, in good agreement with the  
result of Ref.~\cite{est98}, in which the authors quote $(6.3 \pm 1.7)\%$.\\
The calculation of Efremov {\it et al.} is averaged over the quark flavours. 
Since we 
are taking into account only valence contributions, and we are assuming 
isospin symmetry to hold, we have
$H_1^{\perp\,\mbox{\small fav}}/D_1^{u / \pi^+} 
=  H_1^{\perp\,\mbox{\small fav}}/D_1^{\bar d / \pi ^+}$,  
then the value we give can well be compared with the averaged one.
Notice that this evaluation is rather sensitive to the $z$ cut: by lowering 
$z_h(min)$ to $0.01$, for example, the ratio $H_1 ^{\perp a}  /  D_1 ^a$ would 
be reduced to 0.023. On the other hand, choosing a higher value of $z_h(min)$, 
say 0.2 for instance, the ratio $H_1 ^{\perp a}  /  D_1 ^a$ would increase 
to about 15\%. 

Now, a completely analogous calculation can be performed to give an estimate 
of the ratio  $f_{1T} ^{\perp} /  f_1$. Once again we take into account 
only valence quarks contributions in the proton, $u$ and $d$. 
%Because of the much slower encreasing of the distribution function $f_1$ for 
%decreasing values of $x$, no cut is needed to perform the integration. 
By adopting the same cuts as in HERMES and COMPASS experiments, 
$0.02 \leq x \leq 0.4$, we have 
\be
%\frac{f_{1T} ^{\perp u}}{f_1} = 
\left| \frac{\int _{0.02} ^{0.4} \, dx \, 
f_{1T}^{\perp \, u}(x)}{\int _{0.02} ^{0.4} \, dx \,f_1^{u} (x)} \right|
= 0.083\;,
\ee
\be
%\frac{f_{1T} ^{\perp d}}{f_1} = 
\left| \frac{\int _{0.02} ^{0.4} \, dx \, 
f_{1T}^{\perp \, d}(x)}{\int _{0.02} ^{0.4} \, dx \,f_1^{d} (x)} \right|
= 0.072\;.
\ee
Notice that in this case the results are not very sensitive to the $x$ cuts. 
In fact, we would have obtained very similar results by setting the upper 
limit of integration to 1 ($0.11$ and $0.08$ for $u$ and $d$ respectively). 
The same holds when decreasing the lower limit of integration: 
for $x(min)=0.01$, 
for example, we would have had $0.07$ for $u$ and $0.06$ for $d$.\\
Thus, for an average over the flavours (assuming that 
non-valence contributions are negligible) we find that the ratio 
$f_{1T} ^{\perp} /  f_1$ is about $7.7\%$, quite close 
to the result we found for $H_1 ^{\perp a}  /  D_1 ^a $.

We stress that none of the above estimates takes into account effects of 
evolution. Furthermore, just comparing integrated results neglects not only 
several kinematics factors, which we did show in Section III, but also 
forgets about experimental considerations such as azimuthal acceptances, 
etc. 

\vspace{1cm}

\small
\begin{center}
{\bf V. CONCLUSIONS } 
\end{center}

\vspace{3mm}

\normalsize

In this paper we have presented results for some observables in
lepton-proton scattering that 
provide information on time reversal odd distribution and 
fragmentation functions. Far from being precise predictions, our results give 
rough estimates based on experimental data from $p^{\uparrow}p$ single spin 
asymmetries
and on some theoretical prejudice as far as unknown functions are concerned.
For the two extreme possibilities, we have indicated 
the kinematical regions in which these rather exotic 
spin effects are sizeable and we have given their overall size and their 
relations to measurable 
angles ($\phi^l_S$ and $\phi^l_h$). Moreover, we find the ratio between odd 
and standard distribution and fragmentation functions to be of the order of a 
few percent. Thus, if these functions do exist, their presence could be 
experimentally detected.

Experimental input is now needed to deepen our knowledge on spin 
effects in high energy scattering processes.
Once again we want to stress that only a joint effort of cooperation between 
theoretical modelling and experimental measurements will allow us to learn 
more about these soft functions, distributions and fragmentations, which in 
turn will teach us about the non-perturbative phenomena leading to particular
correlations between quark spins and transverse momenta or phenomena
occurring in the hadronization process.

\section*{Acknowledgements}

\noindent
We would like to thank M. Anselmino for many valuable discussions.\\ 
This work is part of the research program of the foundation for the 
Foundamental Research of Matter (FOM) and the TMR program ERB FMRX-CT96-0008.

\end{document}